\documentclass[english]{cccconf}
\usepackage[comma,numbers,square,sort&compress]{natbib}
\usepackage{epstopdf}
\usepackage{graphicx}
\usepackage{grffile}

\begin{document}

\title{Energy balancing using charge/discharge storages control and load forecasts in a renewable-energy-based grids}

\author{
Denis Sidorov\aref{melent,pol},
Qing Tao \aref{csu}, 
	Ildar Muftahov\aref{rj,melent},
Aleksei Zhukov\aref{melent},
        Dmitriy Karamov \aref{melent,pol},
        Aliona Dreglea \aref{melent,pol} and
        Fang Liu \aref{csu}}

% Note: the first argument in the \affiliation command is optional.
% It defines a label for the affiliation which can be used in the \aref
% command. If there is only one affiliation for all authors, then the
% optional argument in the \affiliation command should be suppressed,
% and the \aref command should also be removed after each author in
% \author command, in this case the affiliation will not be numbered.

\affiliation[melent]{Energy Systems Institute of Russian Academy of Sciences,
130 Lermontov Str., Irkutsk 664033, Russia
        \email{\{contact.dns, ildar.mft, zhukovalex13, adreglea\}@gmail.com}}
\affiliation[pol]{Irkutsk National Research Technical University, 
83 Lermontov Str., Irkutsk 664074, Russia}

\affiliation[csu]{School of Automation,
Central South University,  932 Lushan S Rd,  Changsha 410083, China
        \email{csuliufang@csu.edu.cn}}

\affiliation[rj]{ Main Computing Center of JSC Russian
Railways, 25 Mayakovaskogo St., Irkutsk 664005, Russia}

\maketitle

\begin{abstract}
Renewable-energy-based grids development  needs new methods to maintain the balance between the  load and generation using the efficient energy storages models. Most of the available energy storages models do not take into account such  important features as the nonlinear dependence of efficiency on lifetime and changes in capacity over time horizon, the distribution of load between several independent storages. In order to solve these problems the Volterra integral dynamical models are employed. Such models allow to determine the alternating power function for given/forecasted  load and generation datasets. In order to efficiently solve this problem, the load forecasting models were proposed using deep learning and support vector regression models. Forecasting models use various features including average daily temperature, load values with time shift and moving averages.  Effectiveness of the proposed energy balancing method using the state-of-the-art forecasting models is demonstrated on the real datasets of Germany's electric grid.
\end{abstract}


% Please remove or comment out the following line if the footnote is not necessary
\footnotetext{This work was supported in part by the NSFC-RFBR Exchange Program under Grants 61911530132/19-58-53011
and by the  RFBR Grant 18-31-00206. }

\section{Introduction}

Energy industry is on the verge of new changes due to the various renewable energy sources employment. Relations liberalisation among electricity generators, suppliers and consumers poses new fundamental mathematical
problems in the various fields of the modern power engineering.
The high proportion of various renewable energy sources increases the  power generation variability, disrupting the optimal mode of operation of traditional power systems.
The studies of various  energy storages usage, such as pumped power plants \cite{punys2013assessment}, compressed air storages \cite{karellas2014comparison}, rechargeable batteries \cite{dunn2011electrical}, and others, are especially important in such challenging conditions of  power systems upgrade. % \cite{ ali2010electronic}.%l,}. sebastian2012flywhee
Extensive research studies have been conducted on the modeling and optimization of electric energy storages, see e.g. \cite{di2014modelling, makarov2012sizing}.

 R. Dufo-Lopez, J. L. Bernal-Agustin, D. Tsuanyo, E. Dursun \cite{dufo2014comparison, tsuanyo2015modeling, dursun2012comparative} used the method of chronological modeling of the operation of electrochemical energy storages for the determination of the state of charge, voltage, resistance,  current charge/discharge optimal  behaviour.
These studies contribute to a better understanding of the current state of research of power systems using the storages, their technical characteristics, functional limitations and design capabilities of systems using renewable energy sources.
A lot of attention is paid to the analysis of performance, technical characteristics and cost of using different storages including the batteries.
It is to be noted that most of the works employs the linear models of the storages.
Such approaches do not take into account the nonlinear processes of reduction of the available capacity of the energy storages over time and other important features like the state of charge (SOC).

This paper employs a new models based on Volterra integral equations \cite{sidorov2015integral}, which take into account such principal characteristics of storages as capacity, efficiency, number of cycles and discharge / charge rate, load distribution between the available storages.
Proposed integral dynamical model relies on electric load forecast, generation from the renewable energy sources and traditional generation. This article proposes load forecasting models using the state-of-the-art machine learning methods.

Inaccurate forecasts  reduce the quality of grids' management:  forecast errors lead to the need to use an expensive emergency power plants or to purchase a missing capacity from neighboring manufacturers at higher prices; an overestimated prediction leads to an increase in the costs of maintaining the excess reserve capacity.

There are many forecasting methods available in the literature, most of which are based on traditional methods, not considering the latest advances in machine learning and data analysis. This paper proposes an approach based on data analysis techniques using the following machine learning methods:  support vector regression \cite{drucker1997support} (Support Vector Machine based Regression, SVR), deep learning based on recurrent neural networks (Long Short-Term Memory, LSTM; Gated Recurrent Units, GRU) \cite{lstm, gru}, random forest \cite{breiman2001random, zhukov2016random} and gradient boosting \cite{friedman2002stochastic}.

Features engineering is the principal part of the forecasts models construction. It is important to correctly  identify and formalize the  factors  influencing the forecast.
They can be divided into socio-economic and meteorological, which in turn can be cyclical (for example, day of the week), natural (reflecting the natural activity of the technological or natural environment, for example, the heating season or the atmospheric pressure) or random  or sudden changes in weather conditions.

All these factors serve as the input  for the forecast. In order to represent such information it in the most convenient form for the model, the transformations like the principal component analysis or the Hilbert--Huang transform can be used.

Forecasting mathematical models can be classified into the following two groups:
\begin{enumerate}
\item based on traditional statistical and probabilistic methods for analyzing time series; 
\item the artificial intelligence based methods.
\end{enumerate}

The first group of methods includes traditional statistical methods for time series analysis, such as multiparametric regression (hereinafter, linear model, LM), exponential smoothing, autoregression models, moving average, and their modifications, such as ARMA, ARIMA. Such models enjoy good accuracy if the input features do not correlate with each other, and have linear dependence on the target variable. Therefore, they often fail to make an sufficiently accurate forecast for data with complex load dependencies on meteorological and socio-economic factors. In addition, without additional filtering, such methods are unstable with respect to outliers and data errors.
The probabilistic forecasting methods include the statistical gradient method, Bayes models and other. In order to obtain continuous prediction, filters can be used: for predicting stationary processes, the Wiener-Hopf filter is used; for non-stationary processes, a Kalman filter is used. These methods are usually jointly employed with the regression to improve the prediction result and to make it more resistant to incorrect data.

The second approach, which is most often used, uses traditional and the state-of-the-art models of machine learning, optimization methods and aggregation of expert knowledge. Among the models of machine learning, the most commonly used are both the well-known feedforward artificial neural networks (ANN), random forest (RF),  gradient boosting decision trees (GBDTs), support vector regression (SVR), and the deep ANNs. To solve the problems of selecting their parameters, various optimization methods can be used including the genetic algorithms. 

The effectiveness of the artificial intelligence (AI) methods can be explained not only by their ability to approximate the complex hidden dependencies, but AI methods also allow to integrate the various techniques into a single model. For example, expert knowledge can be added to the model as an additional feature.  The significance of variables assessment helps to dynamically find out  the most important features in specific conditions for the forecast of specific parameters.  More details on the different approaches to load forecasting can be found in review \cite{kuster2017electrical}.

The remainder of this paper is organized as follows. Sec. 2  focuses on the mathematical dynamical model of the storages in general settings. Experimental studies are fulfilled in Sec. 3,  where first the dataset is described, then results of load forecasting are given in Sec. 3.1 and results concerning the application of the Volterra model for charge/discharge storage control are provided and discussed in Sec. 3.2. Concluding remarks are given in Sec.4.

\section{Storage Dynamical Model}

In order to efficiently model the storages operation, it is proposed \cite{volt} to use the following nonlinear integral model with constraints
\begin{equation}\label{eq1}
\left\{ \begin{array}{ll}
\mbox{$\int\limits_{0}^{t} K(t,s,x(s)) \; ds = f(t), \,\,\,   0 \leq s\leq t \leq T,\,\, f(0)=0$},\\
%\mbox{$\smash{\displaystyle\max_{i=1,N}}{\;\left(\sum\limits_{j=1}^{i}{x_j(t)}\right) \leqslant v_{max}}$}\\
\mbox{${v(t) = \int\limits_{0}^{t}{x(s) ds}, \; \smash{\displaystyle\max_{t \in [0, T]}}{\;v(t)} \leq v_{max}}$},\\
%\mbox{$E_{min}(t) \leqslant \sum\limits_{i=1}^{N-1}\sum\limits_{j=1}^{N}{x_j(t)}\leqslant E_{max}(t)$}\\
\mbox{$E_{min}(t) \leq \int\limits_{0}^{t}{v(s) ds}\leq E_{max}(t)$},\\
\mbox{$0 < \alpha_1(t)<\alpha_2(t)<$ $\dots$ $< \alpha_{n-1}(t)<t,$}\\
\end{array} \right.
\end{equation}
 where the kernel is represented as follows
 \begin{equation}\label{eq2}
K(t,s,x(s)) = \left\{ \begin{array}{ll}
\mbox{$K_1(t,s)G_1(s,{x(s)}), \,\, t,s \in m_1,$} \\
\mbox{\,\, \dots \,\, \dots \dots \dots } \\
\mbox{$K_n(t,s)G_n(s,{x(s)}), \,\, t,s \in m_n.$} \\
\end{array} \right. 
\end{equation}
Here\\
 $m_i = \{ t, s\,\,  \bigl |\,\, \alpha_{i-1}(t) < s < \alpha_i(t) \};$\\
${ \alpha_0(t)=0,\,\, \alpha_n(t)=t,\, i=\overline{1,n.}};$\\
$\alpha_i(t),$ $f(t) \in \mathcal{C}_{[0,T]}^1,$\\
 $K_i(t,s) \in \mathcal{C}_{[0,T]}^1$  for $t,s \in \overline{m_i};$\\
 $ K_n(t,t) \neq 0$, $K_i(t,s), \, G_i(s,x(s)) \in \mathcal{C}_{[0,T]};$\\
 $\alpha_i(0)=0,$ $0 < \alpha_1(t)<\alpha_2(t)<$ $\dots$ $< \alpha_{n-1}(t)<t;$\\
$\alpha_1(t), \dots , \alpha_{n-1}(t) $  increase for small $\tau,$  $0\leq t \leq \tau;$\\
 $0< \alpha_1^{\prime}(0) \leq$ $\dots$ $\leq \alpha_{n-1}^{\prime}(0)<1. $

The theory of such integral models with piecewise continuous kernels was first proposed in \cite{sidorov2011volterra} and developed in \cite{sidorov2015integral, volt, igu18}. The special linear discrete case of the model of grid-connected storage was employed by 
R. Dufo-L\'opez, see  \cite{dl} and other publications. 
 
 Here $ f(t) $ is the load imbalance 
 \begin{equation}\label{eq3}
f(t)=f_{RES}(t) +f_{gen} - f_{load}(t),
\end{equation}
where $ f_ {gen} (t) $ is the generation of traditional energy sources, $ f_ {RES} $ is the generation of renewable energy sources and $ f_ {load} (t) $ is the predicted load of consumers. 
 
The functions $ \alpha_i (t) $ show the load distribution between $ n $ drives, and $ K_i (t, s) G_i (s, x (s)) $ - the efficiency of each drive, changing under the influence of two factors - the lifetime $ K_i (t, s) $ and the intensity of current use of time $ G_i (s, x (s)) $, depending on the alternating function $ x (s) $. 
In this case, the proposed mathematical model allows us to take into account the nonlinear nature of changes in efficiency depending on the service life and/or the behavior of $ x(s) $.
In the equation (\ref{eq1}), the alternating function of changing the power $ x(s) $ is the desired one. It allows for known $ v_{max} $ (maximum speed of the charge): 
\begin{enumerate}
\item
to determine $ E(t) $ is the storage state of charge under the constraints $ E_ {min} (t) \leq E (t) \leq E_ {max} (t) $ depending on the type of storage; 

\item to determine the minimum total capacity of the storage  to cover the load shortage of consumers; 

\item to calculate the number of  cycles based on behavious of function $ E(t) $ function; 

\item to predict the  lifetime of the storage.

\end{enumerate}

The problem of solution to the equation (\ref{eq1}) with respect to $x(t)$ is the typical inverse problem \cite{sizbook}.
The author's  numerical method from \cite{muftahov2015perturbation} is employed. 
 
 The advantages of the nonlinear mathematical model presented above are as follows:
 \begin{enumerate}
 \item{definition of operating parameters of storages
when various renewable energy sources and storages are used jointly;}
\item{consideration of such characteristics of storages operation as power, charge/discharge rate, maximum number of work cycles, SOC limit;}
\item{minor impact on the computational complexity of the algorithm when using a large number of storages ($ n> 10 $);}
\item{accounting for the nonlinear nature of efficiency changes;}
\item{the ability to flexibly customize the time distribution functions of the load between the drives.}
 \end{enumerate}
 
It is to be noted that for efficient application of the proposed Volterra model it is necessary to construct the accurate forecasts of the generation from RES $ f_ {RES} $ (see e.g. \cite{fang} for the short term wind power forecasting models) and consumer loads $ f_{load} (t) $. This problem is attacked using the state-of-the-art machine learning methods. The next section of the article is focused on the forecasting models and verification of the Voltera model on the real datasets. 
 
\section{Experimental Results}

For testing, the Germany's electrical grid dataset was chosen, since this grid is rather large and has many renewable energy sources and electrical energy storage for leveling the daily heterogeneity of the electrical load graph. At the moment, about 6.7 GW of pumped storage power plants (PSPP) are installed in Germany\footnote{https://www.dena.de/en/topics-projects/energy-systems/flexibility-and-storage/pumped-storage/}.
 
To test the proposed approaches, the publicly available data on the German power grid load, provided by ENTSO-E, for the period from the beginning of 2006 to the end of 2013 was used. It should be noted a significant percentage of renewable energy, which for 2007 was 13.6 and to date, according to the statistical service of the European Union\footnote{http://appsso.eurostat.ec.europa.eu/nui/submitViewTableAction.do} 
almost tripled. We also used data on the generation of electricity from various sources located in Germany\footnote{https://www.energy-charts.de/power.htm}. 
 
However, in the analyzed dataset, the influence of the nonstationary nature of wind turbines is insignificant; this was revealed from the results of a statistical Dickey-Fuller's test for the stationarity. Thus, this series can be considered as stationary. However, it cannot be guaranteed that this situation will remain the same in the future. There are no strict requirements for the period of training the models, then its updating can occur periodically. In the case of non-stationary data, online models can be used, such as OzaBag \cite{oza2005online} and PDSRF \cite{zhukov2016random}.

Most of the load (47\%) according to  German Association of Energy and Water Industries (BDEW) is related to the industrial sector (unlike other countries with a large percentage of wind turbines, where population form most of the load), which may be the cause of the most ordered nature of the electric load. The German population form  26\% of the load, the service sector is 25\%, transport is about 2\%.
 
These loads were supplemented by average daily temperature data obtained by the European Climate Assessment \& Dataset from the weather stations in Hamburg, Munich, Stuttgart, Bochum, as well as an indicator of working days and holidays in Germany.
Also, the following indicators are used for the forecast: current load, day of week, time of day, load a day ago, load value an hour ago, load value a week ago, average load for yesterday, minimum load for yesterday, and exponential moving averages with periods 12, 24, 48, 168 hours.

\begin{figure}[htbp]
	\centering
	\includegraphics[width=1.0\linewidth]{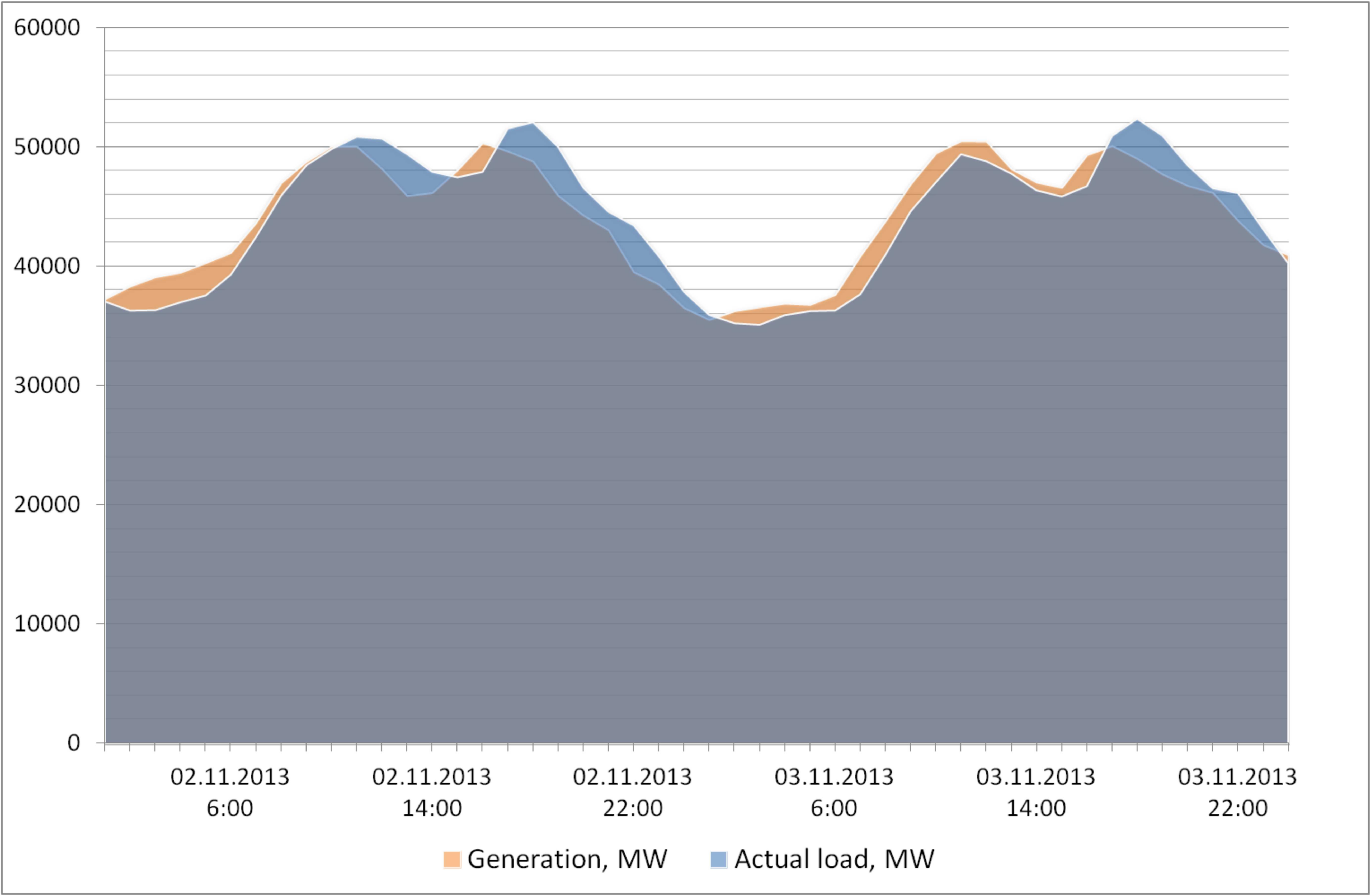}
	\caption{Generation and actual load: from the German grid dataset.}
	\label{fig:GenForecastLoadGermany}
\end{figure}

 The dataset consists of 69713 examples, including data collected from 2006-01-08 to 2013-12-30. Of these, 60953 were used for training and testing predictive models (using block cross-validation), the remaining 8760 for validation.

\subsection{Electric load prediction}
 Six popular models were chosen for prediction:  support vector machine, LSTM, GRU, RF, GBDT and multiparametric linear regression.  Model's parameters are included in the Tab.1. 
 
As can be seen from Table 1, all the tested machine learning methods under consideration have similar errors. Three following metrics were selected:
$ RMSE ~ = ~ \sqrt {\frac{1}{n}\sum_{t = 1}^{n}(x_t-\bar{x_t})^ 2}$ is the root mean square error, mean absolute error $ MAE =\frac{1}{n}\sum_{t = 1}^{n}|x_t-\bar{x_t}|$ and the average absolute error in percent $MAPE=\frac{1}{n}\sum_ {t = 1}^{n}\frac {| x_t- \bar{x_t}|}{\bar{x_t}}*100\% $.
Here $\bar{x_t}$ is the real target value, and $x_t$ is the predicted value.

\begin{table}[!htb]
  \centering
  \caption{Errors Analysis}
  \label{tab1}
  \begin{tabular}{l|l|l|l|l}
    \hhline
    Method          &  Notes & RMSE & MAE & MAPE \% \\ \hline
    SVR   &  RBF kernel & 2472.93 & 1551.38 &  3.41 \\ 
      $\,$ & $C  = 32$  &  $\,$ & $\,$ \\ 
    $\,$ &   $\gamma=0.05515674$  &  $\,$ & $\,$ \\ \hline
    LSTM        & lstm\_1(300)  & 2134.19 & 1236.25 & 2.74 \\ 
    $\,$       & lstm\_2(300) &\,  &\,\\ 
    $\,$      & Dense\_1(16) &\,  &\, \\ \hline
    GRU        & gru\_1(300) & 2114.17 & 1207.96 & 2.68 \\ 
    $\,$    &  gru\_2(300) & \, & \, & \, \\ 
    $\,$         & Dense\_1(16) & \ & \, &  \\  \hline
    RF & {\footnotesize mtry: 4}  &2145.59 & 1667.25& 2.77\\ \hline
    GB & {\footnotesize interaction.depth: 9}\hspace*{-1.2mm} &2144.89 & 1293.78 & 2.89\\ 
      $\,$ & {\footnotesize  shrinkage: 0.1} &\,   &\,  &\,\\  
      $\,$ & {\footnotesize n.minobsinnode: 10}\hspace*{-1.2mm} &\,   &\,  &\,\\ \hline
     LM        &  \, & 4774.54 & 3735.50 & 8.07 \\ \hhline
  \end{tabular}
\end{table}
 
% \begin{center}
%		~~~~~~~~~~~~~~~~~~~~~~~~~~~~~~~~~~~~~~~~{Table 1}\\
%                                {Errors of considered methods.}
%		\begin{tabular}{|l|l|l|l|}
%			\hline
%			& RMSE & MAE & MAPE,\% \\ 
%			\hline
%			%{\color{red}$\blacksquare$}
%			RF & 2145.59 & \textbf{1262.55} & \textbf{2.77} \\ 
%			%{\color{blue}$\blacksquare$} 
%			GB & \textbf{2144.89} & 1293.78 & 2.86 \\ 
%			%{\color{green}$\blacksquare$} 
%			SVM& 2561.60 & 1667.25 & 3.68 \\ 
%			%{\color{gray}$\blacksquare$} 
%			LM & 4774.54 & 3735.50 & 8.03 \\ 
%			\hline
%		\end{tabular}
%%	\end{table}
%\end{center}
%
%\begin{figure}[htbp]
%	\centering
%	\includegraphics[width=1.0\linewidth]{ErrorByTimeNew}
%	\caption{RMS forecast errors for machine learning models (top) and multiparameter linear regression (bottom).}
%	\label{fig:errorsvstime}
%\end{figure}

 Also of interest are errors by days of the week and time of day. It should be noted that all models based on machine learning show similar errors, the largest for Monday, Thursday and the time interval from 7 to 8 hours. This data can be used to further analyze the nature of the load.

It is also necessary to mention the limitations of the proposed predictive model, which include the fact that it does not take into account the structure of the load. For example, since in the shown example most of the industry is occupied, then by including the parameters of the work of large enterprises, it is possible to improve the quality of the forecast. Another factor in the improvement of the model is the consideration of market conditions as additional input parameters.

\subsection{Experiments with integral model of energy storage}

Proposed dynamical model can use nonlinear dependence on time and changes in the power of storage operation in terms of the storage efficiency. The process of determining the efficiency is not a trivial task, since It consists of many factors. For sake of simplicity, in this paper, the constant efficiency of 92\%  is used.

\begin{figure}[htbp]
	\centering
	\includegraphics[width=1.0\linewidth]{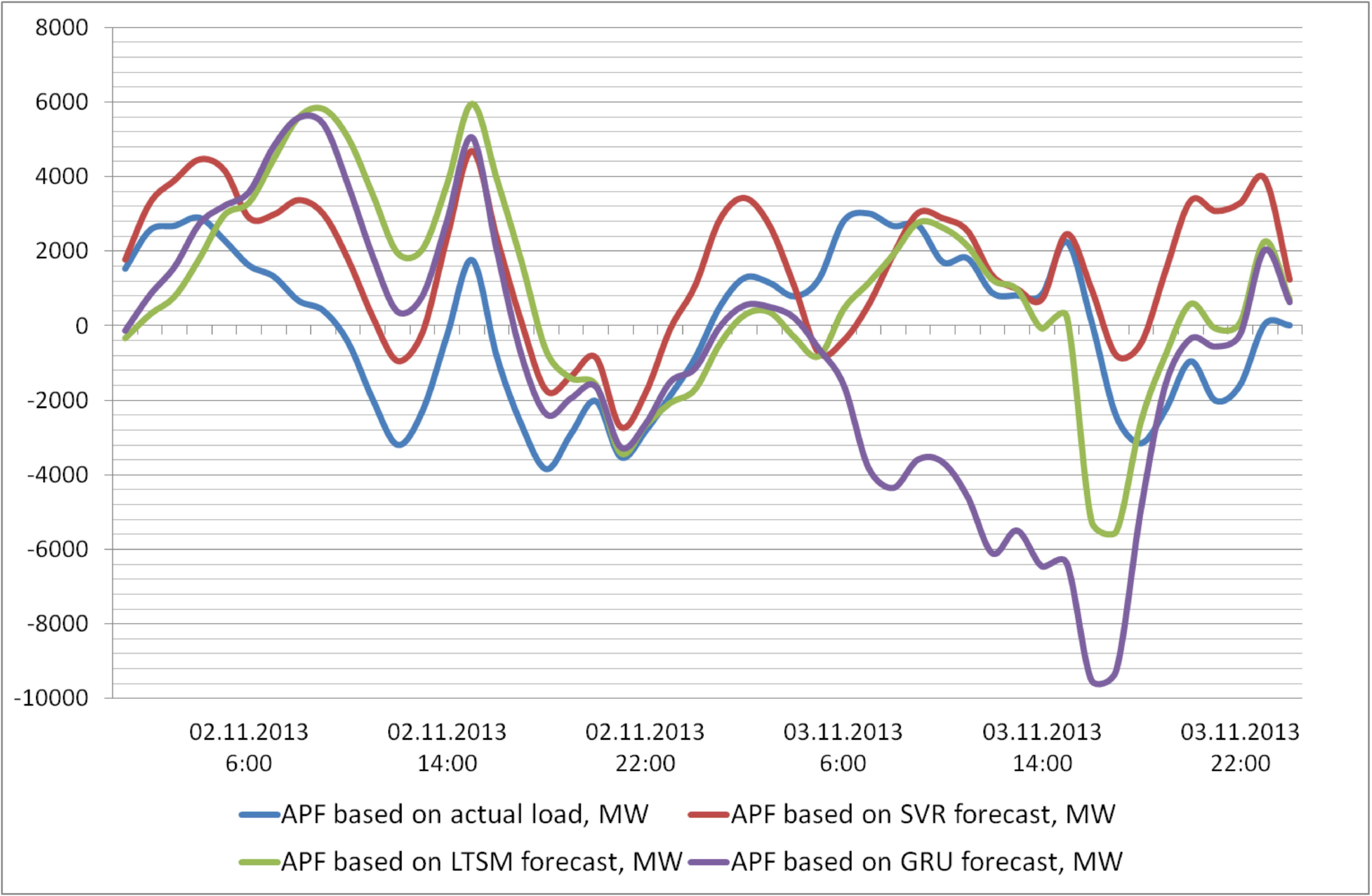}
	\caption{Calculated alternating power functions based on fact  and forecasted loads using deep learning and SVR.}
	\label{fig:BatteryGenGermany}
\end{figure}

 Application of   model (\ref{eq1}) is shown in Fig.\ref{fig:BatteryGenGermany}. It demonstrates the alternating power function based on the actual load and its different forecasts shown in Fig.\ref{fig:GenForecastLoadGermany}. Here the positive values correspond to the process of the storages charging, and negative values corresponds to the generation to cover the load imbalance. It can be noted that the value of the desired function $x(s)$ heavily depends on the forecast accuracy.

\begin{figure}[htbp]
	\centering
	\includegraphics[width=1.0\linewidth]{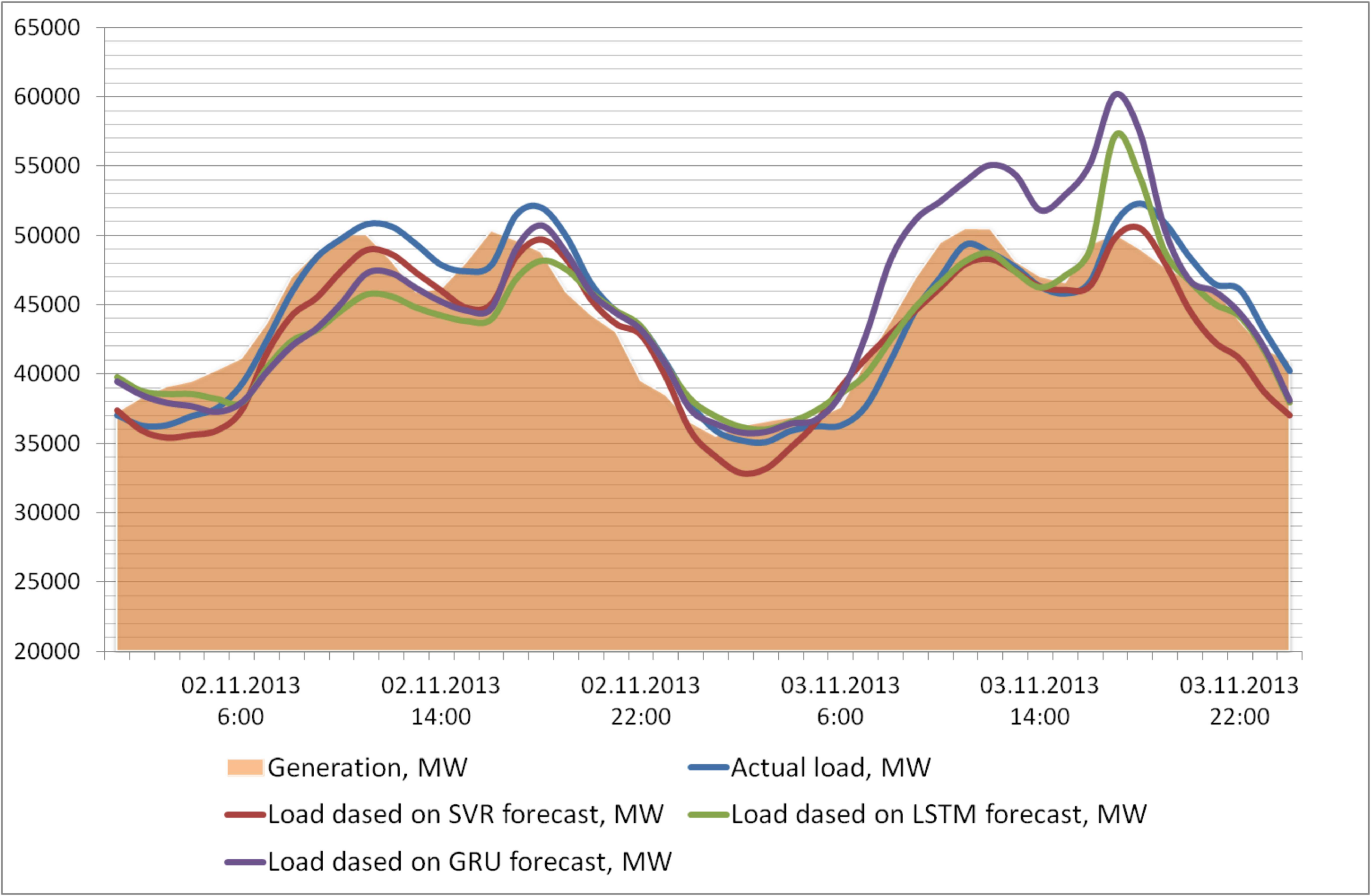}
	\caption{Generation and load forecasting on the data of German grid (deep learning and SVR).}
	\label{fig:GenForecastLoadGermany}
\end{figure}

\begin{figure}[htbp]
	\centering
	\includegraphics[width=1.0\linewidth]{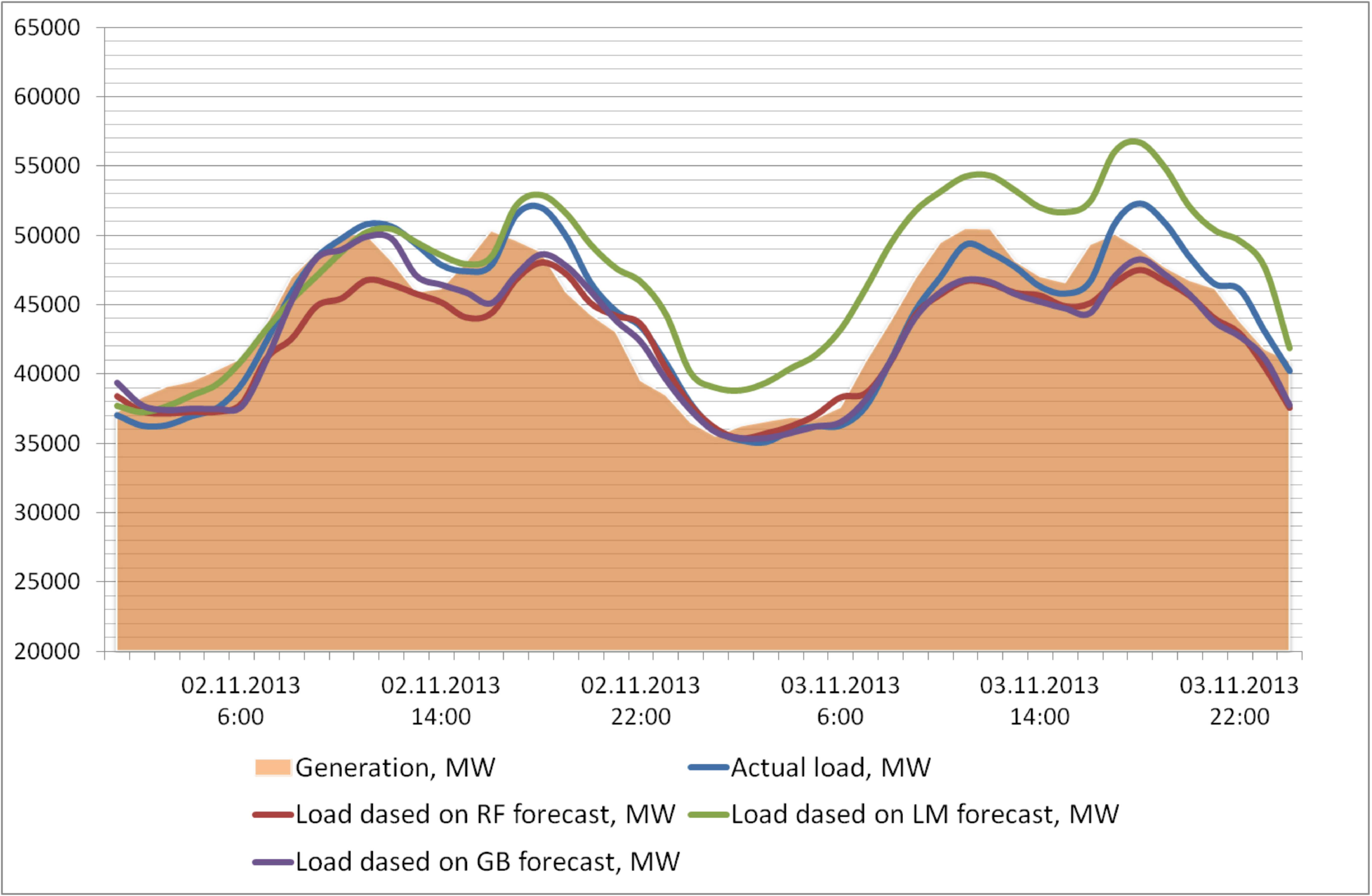}
	\caption{Generation and load forecasting on the data of German grid (RF, GB and LM).}
	\label{fig:GenForecastLoadGermany2}
\end{figure}

\begin{figure}[htbp]
	\centering
	\includegraphics[width=1.0\linewidth]{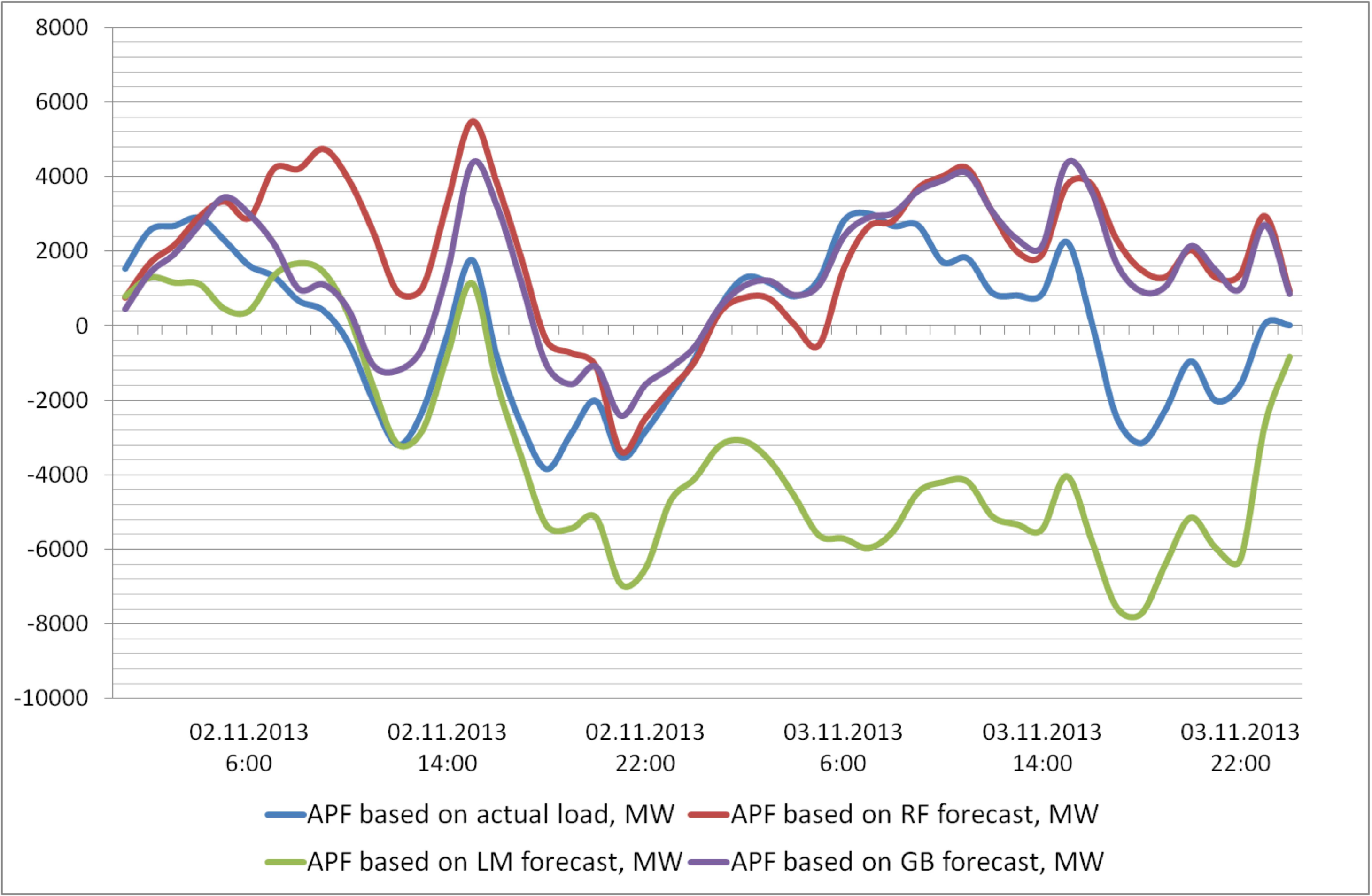}
	\caption{Calculated alternating power functions based on fact  and forecasted  loads using RF, GB and LM.}
	\label{fig:BatteryGenGermany2}
\end{figure}

Fig. \ref{fig:BatteryGenGermany} shows the alternating power functions (APF). As you can see, even a small deviation of the forecast from the actual load gives large differences in the operation of the drive.

As shown by the calculation results (Fig. \ref{fig:BatteryGenGermany}), to cover the imbalance between generation and consumption with storage devices, a minimum of 12 GW of total storage capacity is required.

\section{Conclusion}

This article proposes a new mathematical model of the storages control which allows to take into account the dynamics of efficiency with a nonlinear dependence on the time of use of the storages and APF.
The efficient forecasting models of the electric load for the day ahead are employed, which are based on the deep learning models,
random forest,  gradient boosting decision trees, support vector machine based regression and multiparametric regression. The load forecasting is used as an input parameter for the energy storage model. The proposed models are tested on real dataset of German grid.
The load forecasting models' absolute errors  can be as small as 2.68\%.  Comparison of various  models shows that the best results were achieved using random forest and GRU models. It is to be noted that accuracy of APF is mostly depend on accuracy of the forecast. In our experiments the best accuracy of APF was achieved based on the GRU forecast with MAE = 1272.17. According to the results of calculations with this forecast's accuracy, a full coverage of the imbalance between generation and consumption of energy with storages requires a minimum of 12 GW of total storage capacity.

Our next work will be focused on employment of the self-regularization property  of  considered class of the Volterra models \cite{igu16} and Lavrentiev regularization method \cite{sizbook,sidorov2015integral} in order to cope with inaccurate forecasts of both electric loads and generation.

\end{document}